# A DISTRIBUTED AGENT BASED SYSTEM TO CONTROL AND COORDINATE LARGE SCALE DATA TRANSFERS

**Dobre Ciprian Mihai \*, Ramiro Voicu \*\*, Adrian Muraru \*\*\*, Iosif C. Legrand \*\***

*\* Politehnica University of Bucharest, Romania*
*\*\* California Institute of Technolog, USA*
*\*\*\* European Center for Nuclear Research – CERN, Geneva, Switzerland*

We present a distributed agent based system used to monitor, configure and control complex, large scale data transfers in the Wide Area Network.
The Localhost Information Service Agent (LISA) is a lightweight dynamic service that provides complete system and applications monitoring, is capable to dynamically configure system parameters and can help in optimizing distributed applications.
As part of the MonALISA (Monitoring Agents in A Large Integrated Services Architecture) system, LISA is an end host agent capable to collect any type of monitoring information, to distribute them, and to take actions based on local or global decision units. The system has been used for the Bandwidth Challenge at Supercomputing 2006 to coordinate global large scale data transfers using Fast Data Transfer (FDT) application between hundreds of servers distributed on major Grid sites involved in processing High Energy Physics data for the future Large Hadron Collider experiments.

Keywords: monitoring, distributed systems, agents, peer-to-peer, control plane.

## 1. INTRODUCTION

The High Energy Physics community engaged in CERN's Large Hadron Collider (LHC) is preparing to conduct a new round of experiments to probe the fundamental nature of matter and space-time, and to understand the composition and early history of the universe. The decade-long construction phase of the accelerator and associated experiments is now approaching completion, and the design and development of the computing facilities and software is well-underway. The LHC is expected to begin operations in 2007. The experiments face unprecedented engineering challenges due to the volume and complexity of the experimental data, and the need for collaboration among scientists located around the world. The massive, globally distributed datasets which will be acquired, processed, distributed and analyzed are expected to grow to the 100 Petabyte level and beyond by 2010. Distribution of these datasets will require network speeds of around 10-100 gigabits per second (Gbps) and above. The data volumes are expected to rise to the Exabyte range, and the corresponding network throughputs to the 100 Gbps – 1 Terabit/sec range, by approximately 2015. In response to these challenges, the Grid-based infrastructures developed by collaborations in the US, Europe and Asia such as EGEE, OSG and Grid3 provide massive computing and storage resources. However, efficient use of these resources is tightly connected with the efficient use of the deployed networking equipment.

Even with high-speed networking resources deployed between hosts involved in the LHC experiments, an important problem that needs to be solved is related to End-to-End achievement. During the "High Speed Data Gathering, Distribution and Analysis for Physics Discoveries at the Large Hadron Collider" demonstration at SuperComputing 2006 (SC06)

Bandwidth Challenge (BWC) our agent based solution provided the control plane that helped in the achievement of new records for sustained data transfer.

## 2. THE ARCHITECTURE OF LISA

LISA (Localhost Information Service Agent) is a lightweight framework that can help in optimizing other applications by means of monitoring services. It consists of a number of service agents that can be easily deployed on any workstation, independent of the local architecture or operating system. The LISA agent automatically detects the architecture on which it is deployed and is capable to dynamically load the binary modules necessary to perform monitoring services. LISA is based on Java technologies, any agent being able to operate independent of the operating system platform. The components of the framework are presented in figure 1.

The service agent is composed of a core system and a set of deployable modules. The core system is responsible for managing monitoring modules in the framework. It dynamically discovers available modules and can load or unload them at any time. The modules can be loaded or unloaded without the need of restarting the system and without affecting the rest of the running modules. Another interesting feature of the engine is that is capable of sensing when parameters are updated. When such updates occur the engine is capable of triggering sets of actions such as reloading the modules or updating the running parameters.

The core system is capable of working with remote repositories of modules. From remote locations it can dynamically trigger updates of module. The core system also provides the capability of integrating the agent with another external application using sets of exported API methods. Every module in the system can export such functionality to any external application.

Fig. 2. The information collected by the system monitoring module.

Another aspect of the implementation of the core system is that it is permanently monitoring itself. This capability makes LISA capable of detecting potential problems and choosing from a number of possible solutions to counteract them. For example if a loaded module stops due to some running conditions or programming fault the core system is capable of dynamically restarting it. The core system is even capable of restarting the agent altogether, for example when the Java virtual machine encounters a problem such as the available memory wears out or the running threads come to a deadlock condition.

The LISA Service Agent incorporate a number of monitoring and controlling modules that can be used to completely supervise the end-user system, the monitoring capabilities ranging from the hardware architecture and the running parameters of the workstation to local networking parameters. New modules can be easily constructed and added by users and this capability provides increased monitoring functionality to the system.

One of the already implemented and available modules offers complete system monitoring functionality. This module can monitor environmental information such as the type of the operating system and version, the user under which LISA runs, JVM version, local IP address or, even better, if the address is a private one it can retrieve

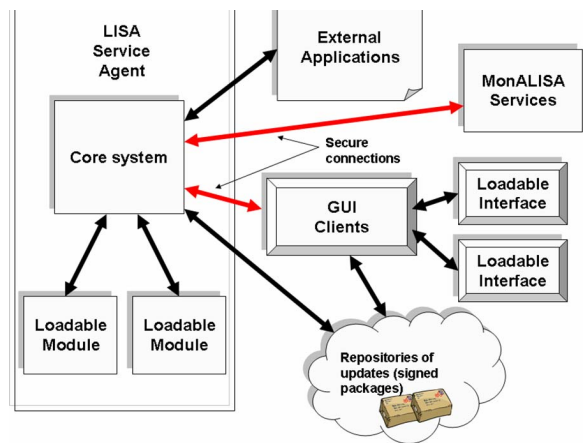

Fig. 1. The components of the framework.

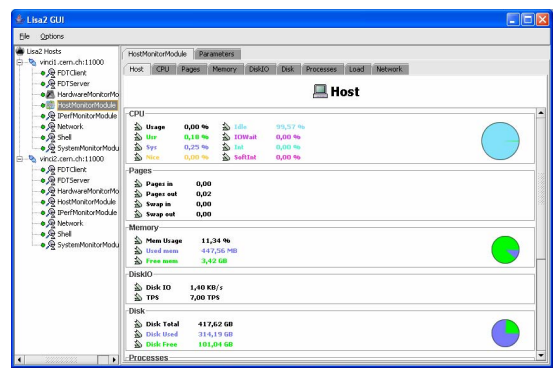

Fig. 3. A global view of the running parameters of the local station.

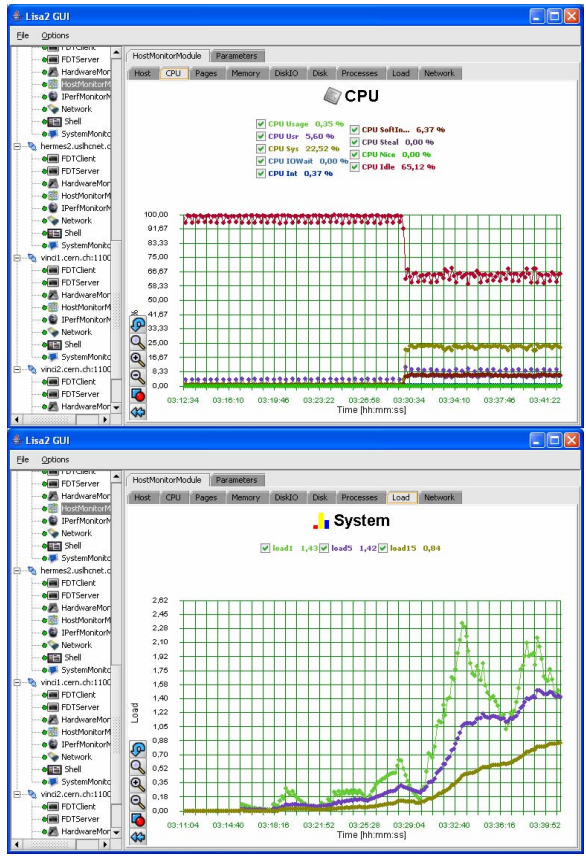

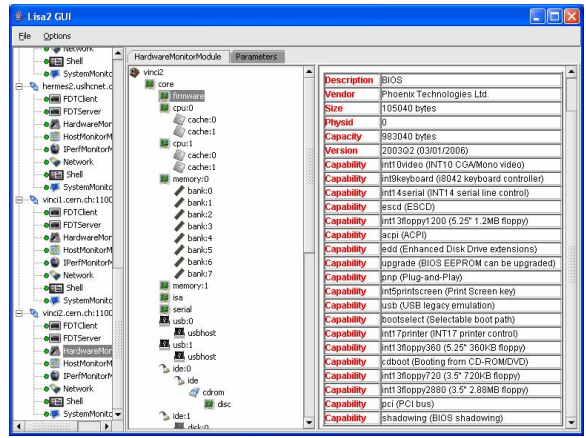

Fig. 5. Hardware monitoring module.

LISA also comes with a module that monitors the hardware configuration of the local workstation. The information collected by this module is independent of the underlying operating system offers. Figure 5 presents an example of the kind of ability presented by this module.

One other module is capable of monitoring and adjusts the networking performances of the local station. This module reads information regarding the network interfaces available in the system such as traffic patterns, addresses and capacities and current settings. Also it monitors specific parameters related to the installed TCP/IP stack. It reads parameters regarding the health of the IP protocol (packets sent and received, bad packets, resent packets, etc.), the health of the TCP protocol (sent and received segments, fragmentation, etc.) and the health of the UDP protocol. This module is also able to adjust on user request some of the parameters that are used for tuning the TCP/IP stack in order to optimize the networking performances of the local workstation. Examples of the capabilities offered by this module are presented in the figure 7.

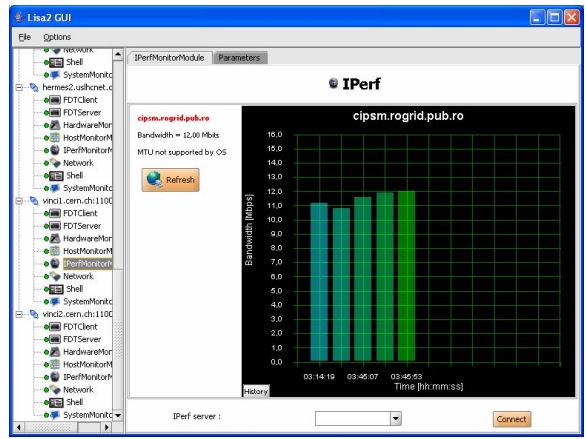

Fig. 6. The module that measure the network performance in terms of available bandwidth.

Fig. 4. Different graphs of values collected by the host monitoring module.

the public address of the access point for the local network, and the AS to which the station belongs. This module is also able to monitor the health of the LISA agent using JMX technology. Figure 2 presents the type of information that this module is capable of gathering from the workstation where LISA is deployed.

One other available monitoring module supervises the status of the workstation health parameters. The monitored information is classified into information regarding the CPU utilization (how much of the CPU is used by the user applications, by the system, how much is idle, etc.), regarding memory utilization (free memory and total available memory in the system, page swapping), regarding disk usage (free disk space and total available disk space), regarding the current load of the system (load1, load5, load15) or the number of running processes and information regarding networking (in and out traffic for each available network interface). This module is dynamically adjusting itself based on the architecture of the system counters of the underlying workstation on which LISA is running. Examples of the capabilities offered by this module are presented in figure 4.

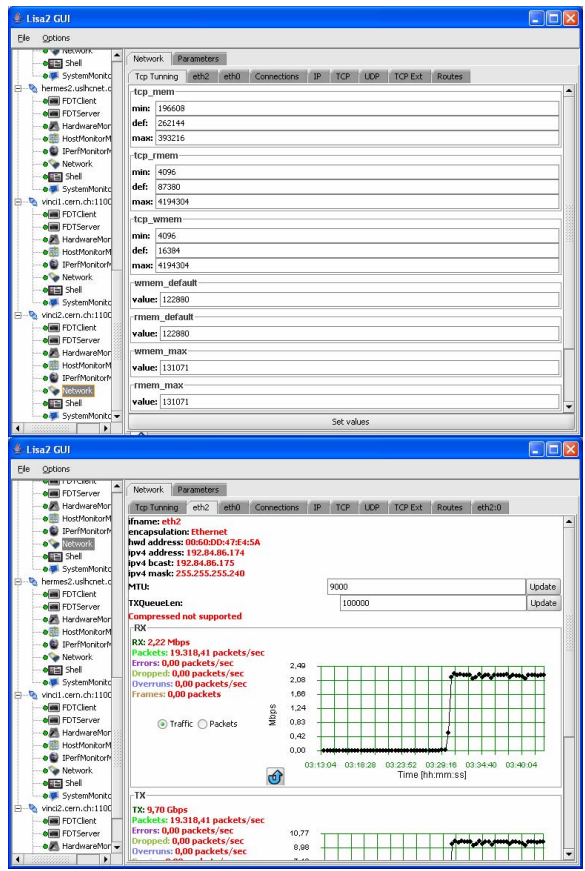

Fig. 7. The network monitoring module and the parameters that can be adjusted.

One other implemented module can be used as a shell interface between any user having correctly defined rights and the local workstation on which the LISA agent is deployed. Using this module the user, meaning the user running the GUI client or an external application, can execute commands and read the returning outputs. And another module is able to test the available networking throughput between the local workstation and any other remote station, using IPERF. Figure 6 presents results obtained using this monitoring module.

All these modules are meant to make LISA an application which best describes the local station in order to help other applications optimizing performances by means of monitoring services. There are actually two defined ways in which an application can integrate the LISA services. Any application can use the API functions that each module is exporting. In this way any application can be tightly coupled with the LISA framework. Another modality is that the application connects itself to the service daemon. In fact this is how LISA is connecting to the higher level MonALISA monitoring services and also this is how the GUI client connects to the LISA services. In both approaches the LISA agent is sending to the registered interfaces the monitoring results and can receives commands to trigger internal actions.

The lightweight monitoring framework also consists of a GUI client which can be used by any user to keep track of the current monitoring results. The GUI client is using Java WebStart technology and because of that it can be executed on any local station.

The GUI client can connect to any number of LISA services using networking communication. In this way the status of the services is monitored from any remote location. When connecting to a remote monitoring agent the user running the GUI client have to authenticate in order to provide a set of rights. Based on this access rights the user can have only the capability to receive results or it can more advance rights such as to modify running parameters or rights to start or stop modules.

Each module loaded in the LISA framework is able to define a graphical interface. The client does not start with any of the graphical interfaces defined by any of the running modules. All these interfaces are dynamically loaded into the GUI client. The interfaces can be loaded from any remote locations, from repositories of modules or even can be loaded from the agent itself where a particular module is running. When the user connects to a running LISA agent it will not receive from the beginning all monitored results. When the user becomes interested in a particular module (by clicking on that particular module) the dynamic interface is loaded in the GUI system and the client starts receiving monitoring results.

The client is able to preserve the list of known LISA agents from one run to another. This combined with the aspect that the GUI client is able to remember certain repeated actions means that the system is designed to be user-friendly.

Because LISA is a part of MonALISA there is another module which sends the monitored values back to one or more MonALISA service. For that this modules uses the ApMON module provided in the MonALISA framework. The MonALISA can help in the way that the monitored result values can be processed at a higher level using the MonALISA inner functions.

## 3. FAST DATA TRANSFER ARCHITECTURE

FDT is an Application for Efficient Data Transfers which is capable of reading and writing at disk speed over wide area networks (with standard TCP). It is written in Java, runs an all major platforms and it is easy to use. FDT is based on an asynchronous, flexible multithreaded system and is using the capabilities of the Java NIO libraries. FDT performs the streaming of a dataset (list of files) continuously, using a managed pool of buffers through one or more TCP sockets. It uses independent threads to read and write on each physical device and transfers data in parallel on multiple TCP streams, when necessary. It uses appropriate-sized buffers for disk I/O and for the network FDT restores the files from buffers asynchronously and resumes a file transfer session without loss, when needed.

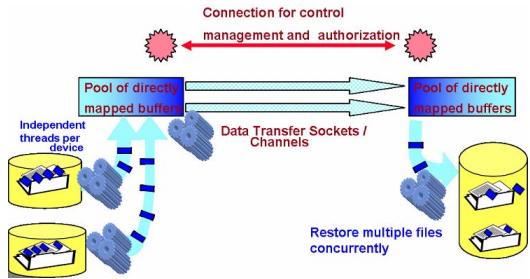

Fig. 8. The architecture of the FDT application.

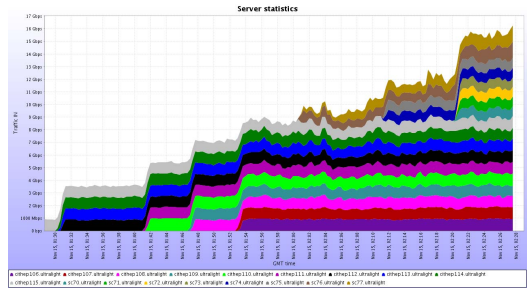

Fig. 9. Results obtained by consecutively adding one more data stream to the overall network transfer flux.

FDT can be used to stream a large set of files across the network, so that a large dataset composed of thousands of files can be sent or received at full speed, without the network transfer restarting between files.

The architecture of the Fast Data Transfer application is described in figure 8.

## 4. LISA AT SC06 BANDWIDTH CHALLENGE

The future of Grid computing lies in the ability to collaborate and transfer petabytes of information around the world. Experiments such as CMS and Atlas are currently constructing detectors for CERN's Large Hadron Collider (LHC) which will unite thousands of physicists worldwide. In order to support such a global project, worldwide interconnections between large databases, mainframes servers, clusters and terminals are required to provide access, processing and analysis of the petabyte sized datasets that experiments such as LHC will provide. The interconnections between machines need to be capable of transferring the traffic that the new datasets will generate.

To keep this flow of information free from disruption, it is not only the servers and clusters that have to be kept operational and optimal, but also the underlying networks that feed the processing farms. It is therefore critical to understand the performance of the underlying networks in order to plan, prepare and better utilize the network. The Fast Data Transfer (FDT) application was designed to meet this scope, to optimize long-range network transfers. During the "High Speed Data Gathering, Distribution and Analysis for Physics Discoveries at the Large Hadron Collider" demonstration at SuperComputing 2006 (SC06) Bandwidth Challenge (BWC), an international team of physicists, computer scientists, and network engineers led by the California Institute of Technology, CERN, and the University of Michigan joined forces to set new records for sustained data transfer between storage systems. Using FDT for its networking support and LISA as the control plane of the experiment the team achieved a peak throughput of 17.77 Gbps between clusters of servers located at the show floor and at Caltech.

Following the rules set for the SC06 Bandwidth Challenge, the team used a single 10-Gbps link that carried data in both directions. FDT provided sustained total throughput of around 17 Gbps for disk to disk transfer using 10 pairs of small servers (each having 4 SATA HD configured in software raid0 and 1Gb/s network interface) in both directions. The official results obtained are presented in figure 10.

During this event LISA played an important role in establishing the record. Through continuous monitoring and supervising tasks LISA provided the control plane for the challenge. This is a perfect example of how LISA can help optimize other applications.

The networking module was used to continuously supervise and diagnose the state of the networking parameters of the servers deployed for the challenge, both on the conference flour and at Caltech. Problems related to the TCP/IP stack were solved quicker then ever because of the way LISA can report the state of various parameters. The network monitoring module made possible to dynamically adjust important configuration parameters remotely and from a single central point of command. The deployed LISA agents collected the global state of the distributed computing nodes and reported in real time the status of the system to the connected clients.

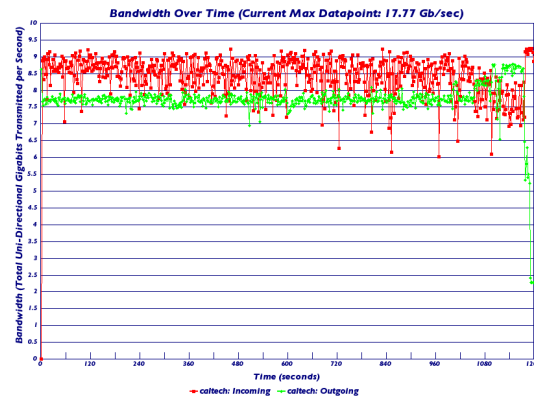

Fig.10. The official results of the Caltech's team at Bandwidth Challenge.

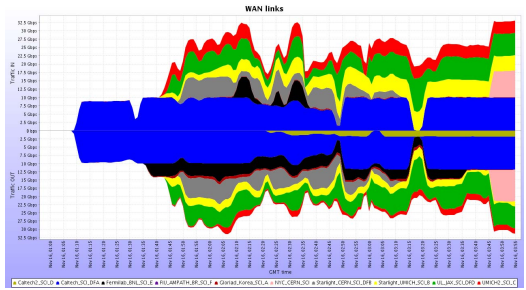

Fig. 11. LISA controlled the network flow obtained by sending data to multiple destinations.

For this event all monitoring module provided precious information about the state of the system. The parameters reported by the host monitoring module provided good insights of the internal state of the computing nodes between which the transfers took placed. The bash module was used to execute diverse commands on the remote nodes in a simple and intuitive way.

LISA also provided one module that was used in close relation with the FDT application. This module reported the state of the application and information pertaining to the network transfers in progress. LISA provided feedback in real-time of the status of the controlled system. LISA was used to dynamically adjust the running parameters of the FDT application, was used to restart the running networking transfers or to change the files being transferred. And LISA performed everything in real time. Operations such as kernel parameters retrieval or asymmetric MTU detection performed by LISA also proved to be crucial in establishing the new records.

The high-energy physics team also carried out several other demonstrations, making good use of the ten wide-area network links connected to the Caltech/CERN booth. Some results are represented in figures 9 and 11. The challenge demonstrated a realistic, worldwide deployment for distributed, data-intensive applications capable to effectively use and coordinate the network resources. In the overall efforts for obtaining the results an important role was played by LISA.

## 5. LISA AS PART OF THE EVO SYSTEM

Another example of the way LISA can help other applications in terms of optimization is the relationship between the framework and the EVO videoconference system. As mentioned in (Dobre, *et al.*, 2004), LISA was made part of the functioning EVO system, assisting the EVO peers to dynamically detect the best reflectors to which to connect to. The best reflectors are read from MonALISA repositories and are updated in time so that the choosing is performed only from the available reflectors. Then the best reflectors are chosen based on their network location (Network domain, AS domain, Country, Continent) and on their current load values, number of currently connected clients, current network traffic. This insures a load balancing order in choosing the reflectors. Based on this from all the available reflectors the module that chooses some and performs RTT values measurements. In the end the module is able to inform some application (e.g. VRVS client) to reconnect to the most appropriate module if network conditions change and the quality is affected.

## 6. CONCLUSION AND FUTURE PLANS

The distributed agent system based on LISA provides the functionality to monitor and dynamically control large scale systems. It was been successfully used to coordinate and optimize large scale data transfers. It is currently used in the EVO videoconference system to optimize the connectivity and performance from the user's perspective. LISA can be used together with the MonALISA framework to provide optimized workflows in distributed grid systems. We are working to provide support for dynamical optical path allocation between end systems.